\documentclass [12pt]{aastex}
\usepackage{graphics,graphicx}

\def \yskip{\penalty-50\vskip3pt plus 3pt minus 2pt}
\def \reference{\par \yskip \noindent \hangindent .4in \hangafter 1}
\def \abc#1#2#3#4 {\reference#1, {\sl#2}, {\bf#3}, #4}
\def \blank {\lower 5pt\hbox to 0.75in{\hrulefill}}

\def \lesssim{\mathrel{<\kern-1.0em\lower0.9ex\hbox{$\sim$}}}
\def \gtrsim{\mathrel{>\kern-1.0em\lower0.9ex\hbox{$\sim$}}}
\def \BD{BD~+30$^\circ$3639}

\begin{document}

\title{On the Asymmetries of Extended X-ray Emission from Planetary Nebulae}

\author{Joel H. Kastner$^1$, Jingqiang Li$^1$, Saku D. Vrtilek$^2$, 
Ian Gatley$^1$, K. M. Merrill$^3$, and Noam Soker$^4$ \\
\small
1. Chester F. Carlson Center for Imaging Science, Rochester Institute of 
Technology, 54 Lomb Memorial Dr., Rochester, NY 14623; JHK's
email: jhk@cis.rit.edu \\
2. Harvard-Smithsonian Center for Astrophysics, Cambridge, MA
02138; saku@cfa.harvard.edu \\
3. National Optical Astronomy Observatories, Tucson, AZ; merrill@noao.edu \\
4. Department of Physics, Oranim, Tivon 36006, 
ISRAEL; soker@physics.technion.ac.il}

\begin{abstract}
Chandra X-ray Observatory (CXO) images have revealed that
the X-ray emitting regions of the molecule-rich young
planetary nebulae (PNs) \BD\ and NGC 7027 are much more
asymmetric than their optical nebulosities. To evaluate the
potential origins of these X-ray asymmetries, we analyze
X-ray images of \BD, NGC 7027, and another
planetary nebula resolved by CXO, NGC 6543, within specific
energy bands.  Image resolution has been optimized by
sub-pixel repositioning of individual X-ray events. The
resulting subarcsecond-resolution images reveal that the
soft ($E < 0.7$ keV) X-ray emission from \BD\ is more
uniform than the harder emission, which is largely confined
to the eastern rim of the optical nebula. In contrast, soft
X-rays from NGC 7027 are highly localized and this PN is
more axially symmetric in harder emission. The 
broad-band X-ray morphologies of \BD\ and NGC 7027 are 
highly anticorrelated with their distributions of visual
extinction, as determined from high-resolution, space- and
ground-based optical and infrared imaging. Hence, it
is likely that the observed X-ray asymmetries of these
nebulae are due in large part to the effects of nonuniform 
intranebular extinction. However, the energy-dependent X-ray
structures in both nebulae and in NGC 6543 --- which is by
far the least dusty and molecule-rich of the three PNs, and
displays very uniform intranebular extinction --- suggests
that other mechanisms, such as the action of collimated
outflows and heat conduction, are also important in
determining the detailed X-ray morphologies of young
planetary nebulae.
\end{abstract}


\section{Introduction}

Models of the formation of PNs have long
predicted that these objects, representing very late stages
in the deaths of intermediate-mass (1-8 $M_\odot$) stars,
should emit X-rays. Such emission 
should arise in shocks at the interface between an active 
wind from the PN core (or its companion)
and material ejected when the progenitor 
was on the asymptotic giant branch (AGB). Thus, extended
X-ray emission, if present, likely traces
the very processes responsible for sculpting PNs 
(for recent discussions of PN shaping mechanisms see, e.g., Frank 1999;
Gardiner \& Frank 2001; Kastner, Soker, \& Rappaport 2000a; and
Soker \& Rappaport 2000).   

The Chandra X-ray Observatory (CXO), with its unprecedented
spatial resolution, has now 
provided the first conclusive evidence of such 
extended X-ray emission from nebular gas, in the form of
striking X-ray imagery of the young PNs BD $+30^\circ 3639$
(Kastner et al.\ 2000b, hereafter KSVD), NGC 7027 (Kastner,
Vrtilek, \& Soker 2001, hereafter KVS), and NGC 6543 (Chu et
al.\ 2001). Observations of the PN NGC 7009 by XMM-Newton,
which combines high sensitivity with good 
spatial and spectral resolution,
also reveal marginally extended X-ray emitting gas (Guerrero, Chu,
\& Gruendl 2002). The PN NGC 7293 (the Helix), known to
exhibit relatively hard X-ray emission, does not display evidence for
extended emission in CXO imaging (Guerrero et al.\
2001); instead, this PN and NGC 6543 contain point-like
X-ray sources with temperatures of a few times $10^6$ K,
possibly due to magnetic activity on companions to their central stars
(Guerrero et al.\ 2001; Gruendl et al.\ 2001; Soker \& Kastner 2002).

While revealing the diffuse nature of the X-ray emission
from some PNs, these first CXO and XMM-Newton observations
of PNs are notable and surprising in several respects.  Of
particular interest is the result that the X-ray
morphologies of the young PNs BD $+30^\circ 3639$ and NGC
7027 are decidely asymmetric, much more so than their
optical nebulosities. Although the extended emission
detected in the CXO images of these PNs (and NGC 6543 and
7009) underscores the importance of strong shocks in shaping
planetaries, the asymmetric structures observed by CXO in BD
$+30^\circ 3639$ and NGC 7027 cannot be easily explained in
terms of ``fossil,'' spherical AGB envelopes acted on by isotropic
white dwarf winds. However, as BD $+30^\circ 3639$ and NGC 7027 are also
among the youngest, dustiest, and most molecule-rich of
known PNs, it is important to assess the extent to which
their observed X-ray asymmetries represent intrinsic
plasma density or temperature structure inhomogeneities
rather than, e.g., the effects of intervening X-ray absorption.  

In this paper, we explore the origin of asymmetries of
extended X-ray emission from PNs (in a companion paper,
Soker \& Kastner 2003, we examine models to explain the
observed X-ray luminosities and temperatures of PNs).  In
particular, we consider the role of nonuniform intranebular
extinction in determining the X-ray morphologies of \BD, NGC
7027, and NGC 6543. In doing so, we take advantage of the
subarcsecond imaging potential of CXO and the energy
resolution of CXO's Advanced CCD Imaging Spectrometer
(ACIS). In \S 2 we present archival optical and
near-infrared images of \BD\ and NGC 7027 that are useful in
deducing their distributions of intranebular extinction, and
we summarize the application of a sub-pixel image
restoration technique to CXO images of these nebulae and NGC
6543. Results, including presentation of super-resolved,
energy-band images of all three PNs and comparisons
of their X-ray surface brightnesses and spatial distribution of
intranebular extinction (as inferred from the
optical/near-IR data), are presented in \S 3. In \S 4, we discuss our
main findings, and \S 5 contains a summary.

\section{Observations and data processing}

\subsection{Optical and Infrared}

Images of \BD\   in the transitions of H$\alpha$
(0.6563 $\mu$m) and P$\alpha$ (1.87  $\mu$m) utilized in the
present analysis (\S 3) were obtained
with the {\it Hubble Space Telescope} using
the WFPC2 and NICMOS cameras, respectively. These images 
appear in Fig.\ 1. The H$\alpha$ and P$\alpha$ images
were first presented in Sahai \& Trauger (1998) and Latter et al.\ (2000a), 
respectively; in addition, HST narrow-band images covering many diagnostic
emission lines were presented in Arnaud et al.\ (1996) and
Harrington et al.\ (1997). The reader is referred to these
papers for details concerning the structure and physical conditions
of \BD\   as ascertained from
HST imaging. As is readily apparent from Fig.\ 1, however,
the inner nebula consists of a bright 
elliptical shell with major and minor axes of $\sim4''$ and
$\sim3''$, respectively (this inner nebula is surrounded by a much larger,
fainter H$\alpha$ halo; Sahai \& Trauger 1998). There 
is a strong gradient in extinction across this region of the nebula, with
many conspicuous knots and clumps of denser gas apparent in
projection against the bright elliptical shell (Harrington et al.\ 1997).

Images of NGC 7027 in the transitions of Br$\gamma$
(2.16 $\mu$m) and Br$\alpha$ (4.05 $\mu$m) utilized in the
present analysis (\S 3) were obtained with the National
Optical Astronomy Observatories\footnote{National 
Optical Astronomy Observatory is operated by the Association of 
Universities for Research in Astronomy, Inc. (AURA), under cooperative 
agreement with the National Science Foundation.} 4 m
telescope and Cryogenic Optical Bench (COB) at Kitt
Peak National Observatory in 1995 September (for details
concerning COB imaging on the 4 m, see Weintraub et al.\
1996). These images appear in Fig.\ 2. Like \BD, NGC
7027 displays a bright, elliptical shell, with a fainter
surrounding halo; the major and minor axes of the shell are
$\sim10''$ and $\sim7''$, respectively. While
extensive near-infrared imagery of NGC 7027 has been
obtained in the 2--3 $\mu$m wavelength range (e.g., Graham et al.\
1993; Kastner et al.\ 1994, 1996; Latter et al.\ 2000b; Cox
et al.\ 2002), the Br$\alpha$ image presented here represents the
longest wavelength image of the nebula presently
available at $\sim1''$ resolution. As we demonstrate in \S 4,
this image is valuable for examining, at high
spatial resolution, highly obscured regions of the nebula.

\subsection{X-ray}

BD $+30^\circ 3639$ and NGC 7027 were observed by CXO
in 2000 March and 2000 June, with net integration times of
18.8 ks and 18.2 ks, respectively. Both PNs were imaged
with the central back-illuminated CCD in the ACIS array (ACIS-S3). 
These observations were
first presented in KSVD and KVS. NGC 6543 was observed
with ACIS-S3 for 46.0 ks in  2000 May (Chu et al.\ 2001). In this paper, we
make use of these CXO/ACIS-S3 event data as
reprocessed by the Chandra X-ray Center (CXC) in 2000
December for BD $+30^\circ 3639$ and in 2001 January for NGC
7027 and NGC 6543 (Figs.\ 3, 4, \& 5, respectively). 

\subsubsection{Event position refinement}

The FWHM of the core of the point spread function (PSF) of
the CXO High Resolution Mirror Assembly (HRMA) is nearly
equivalent to the ACIS-S3 pixel size, $0.49''$.\footnote{See
Chandra X-ray Observatory Proposer's Guide, V. 4
(http://cxc.harvard.edu/udocs/docs/POG/MPOG/index.html), \S
5.4.} Because (1) the telescope position is dithered across
the source during an observation and is known to an accuracy
of better than $0.3'',$\footnote{Chandra X-ray Observatory
Proposer's Guide, V. 4, \S 4.2.3.} (2) the charge cloud
size generated by an incident photon is much smaller than
the ACIS pixel size (Tsunemi et al.\ 2001), and (3) the ACIS
flight software records the distibution of charge among
pixels for each candidate X-ray event (encoded as
``FLTGRADE,'' the bitmap value of pixels above the ``split
threshold''\footnote{Chandra X-ray Observatory Proposer's
Guide, V. 4, \S 6.3.}), it is possible (in principle) to
reconstruct the position of each detected X-ray to sub-pixel
accuracy (e.g., Tsunemi et al.). We have thus applied event
position corrections to the reprocessed data obtained for BD
$+30^\circ 3639$, NGC 7027, and NGC 6543, following the
general method outlined in Tsunemi et al., as implemented,
refined, and tested by Li et al.\ (2002). That is, we move
the position of a photon from the default (pixel center) to
its inferred landing location in detector coordinates,
according to the event grade (see below). We then 
use the telescope pointing and spacecraft roll angle history for
the observation to project the revised event locations to sky
coordinates. 

The Tsunemi et al.\ (2001) method relies on rejection of all
events except those with charge distributed among three or
four pixels; these ``corner split'' events have better
sub-pixel accuracy in both directions (horizontal and
vertical) than single events and in one direction than
two-pixel events, thereby maximizing the potential
improvement in spatial resolution. However, corner splits
constitute only about 10\% of the total number of events for
a typical ACIS-S3 observation of a soft source (e.g., Table 1). Thus, to boost
the signal-to-noise ratio, Li et al.\ (2002) include
single-pixel as well as two-pixel events for which the
charge is split either horizontally or vertically. As we
demonstrate below, the improvement in image quality remains
significant for this more inclusive FLTGRADE selection. In
calculating new photon landing positions in detector
coordinates, Li et al.\ assume single pixel events
correspond to photons absorbed at the event pixel center,
2-pixel vertical or horizontal split event photons land at
the centers of the split boundaries, and the corner split
event photons land at the split corners. Before applying
this event position relocation algorithm, position
randomization within an ACIS pixel (performed as part of the
standard CXC event processing pipeline) must be reversed.

In Table 1 we summarize the event FLTGRADE distributions for
\BD, NGC 7027, and NGC 6543. In compiling these statistics
we have selected those events with nominal energies 
$<3.0$ keV (as determined from calibrations
produced by the standard CXC ACIS
processing pipeline) that lie within the \BD, NGC 7027, and
NGC 6543 source regions as defined in KVS, KSVD, and Chu et
al.\ 2001, respectively. It can be seen that our event
selection criteria retain $\ge 95$\% of events in the source
regions (the selected events necessarily include a small
percentage of background events).

\section{Results}

\subsection{Sub-pixel image reconstruction}

Results of application of subpixel event relocations to the
PN event data are displayed in Figs.\ 3, 4, and 5. In each
Figure, the original image --- i.e., the image obtained by
spatially binning the Level 1 event data provided via
standard event processing by the CXC, after filtering on
energy and FLTGRADE (\S 2.2.1) --- is presented in the
left-hand panel. Images constructed from 
``unrandomized'' event positions are displayed in the center
panels of each Figure; in both these unrandomized images and
in the ``original'' images constructed from Level 1 events,
the pixel size is set to the intrinsic pixel size of ACIS,
$0.49''$. Finally, the right-hand panels of each Figure
display images constructed after relocating
events according to the subpixel event repositioning algorithm. 
These images were constructed for a pixel size of
$0.25''$, which fully samples the HRMA PSF at
the energy range characteristic of the three PNs (0.3 keV to 3.0 keV).

The comparisons between ``original,'' ``unrandomized,'' and
``event relocated'' images in Figs.\ 3, 4, and 5 illustrate
the superior spatial resolution afforded by
subpixel event repositioning. Perhaps the best example of
such an improvement is the sharply decreased FWHM of the
central point source in NGC 6543 (compare the left- and
right-hand panels of Fig.\ 5): before correction we measure
a FWHM of $1.12''$, and after correction we measure a FWHM
of $0.80''$. Similarly, after
event relocation is applied, the X-ray outline of \BD\ and
the bright region of emission at its eastern edge appear
sharper (Fig.\ 3, right). The number of counts per pixel is
quite small in the reconstructed image of NGC 7027, so some
of the structure in this image is due to small number
statistics. Nevertheless, features in the brighter regions
of these images (e.g., the compact, bright region located northwest of
the position of the central star) appear to correspond to
features seen in optical and near-infrared images of this PN
(\S\S 3.2, 3.3).

The absolute astrometry of the 
reprocessed BD $+30^\circ 3639$ and NGC 7027 data is
improved over that available to KSVD and KVS, due to refined aspect
solutions. Hence we can use the reconstructed images of
these nebulae to re-examine the correspondence of their
X-ray and optical emitting regions. For \BD, 
the relative alignment of the P$\alpha$ image and the broad-band
(0.3-3.0 keV) reconstructed image constructed from the reprocessed
CXO event data agrees to within $\sim0.5''$, based on 
comparison of the outlying CXO image contours with the
elliptical shell seen in P$\alpha$ (Fig.\ 6). 
For NGC 7027, we have adjusted the position of the X-ray image by
$\sim1''$, to better align the positions of broad-band peak
X-ray surface brightness with the bright elliptical
shell in the Br$\alpha$ image (Fig.\ 7). 
The resulting overlays of reconstructed
broad-band X-ray images on infrared images (Figs.\ 6, 7)
indicate that (a) the central stars of both PNs are
confirmed to lie very near the center of their X-ray
nebulosities, and (b) unlike both NGC 7293 and NGC 6543 (Guerrero et al.\
2001), neither BD $+30^\circ 3639$ nor NGC 7027 clearly
contains an X-ray-bright central star (a possibility left
open for NGC 7027 by the preliminary analysis of KVS).

\subsection{Energy-resolved images}

In Figs.\ 8, 9, and 10 we present, respectively, 
broad-band and energy-resolved images of \BD, NGC
7027, and NGC 6543. The energy-resolved images span the energy 
ranges $E \le 0.7$ keV (hereafter ``soft-band''), 0.7 keV $< E
\le 1.2$ keV (``medium-band''), and 1.2
keV $< E \le 3.0$ keV (``hard-band''). Like the broad-band
images, the energy-resolved images 
also have been reconstructed at
sub-pixel resolution, with $0.25''$ pixels, via the
algorithm described in the previous 
section. Our analysis of the sub-pixel event positioning
algorithm suggests that the results of the algorithm are
robust, regardless of the count rate of the source, since
the algorithm operates on one photon at a time (Li et
al.\ 2002). Hence, the subpixel reconstructions are reliable,
despite the small number of counts (and, hence, low signal-to-noise
ratio) in certain energy-resolved
images (e.g., hard-band images of \BD\ and NGC 6543, and all
images of NGC 7027).

The results demonstrate that the diffuse X-ray emission
morphologies of all three nebulae are highly
energy-dependent within the 0.3-3.0 keV band. In all four
images of \BD\ it is apparent 
that the X-ray emission is brightest toward its eastern rim
(Fig.\ 8); however, this asymmetry is much more profound in the
medium- and hard-band images than in the soft-band image. In
the soft-band image, the peak of the surface brightness of
\BD\ lies close to (though $\sim1''$ east of) the position
of the central star, while in the medium- and hard-band
images, the nebula is much less centrally peaked. In these
images, the (clumpy) emission from just inside the rims of
the optical nebula appears generally stronger than the
emission from the core region, and there is a compact bright region
at the eastern rim.

In contrast to \BD, the soft-band emission from NGC 7027 is highly
localized (Fig.\ 9). This emission is entirely confined to a region in
the northwest of the nebula that appears as a ``hole'' in
high-resolution optical images (see, e.g., Fig.\ 1 of
KVS). Emission from the X-ray-emitting ``lobe'' projected northwest of
the central star --- a direction corresponding to
blueshifted atomic and molecular gas 
(Cox et al.\ 2002) --- dominates the
nebula in both the soft- and medium-band images. In
the hard-band image the emission is more balanced between
the northwest (forward-facing) and southeast
(rearward-facing) X-ray lobes, and
the northwest X-ray lobe has a smaller opening angle
than in the medium-band image. The outline of NGC 7027 in
this hard-band image appears more or less axisymmetric or, perhaps,
point-symmetric. 

Fig.\ 10 demonstrates that all diffuse X-ray emission
detected from NGC 6543 emerges at energies $< 1.2$ keV; only
the central star is detected in the hard-band image. This is
consistent with the spectral analysis in Chu et al.\ (2001)
and Guerrero et al.\ (2001), which indicates that the
diffuse emission has a characteristic temperature $\sim10^6$
K, while the unresolved central source is somewhat hotter,
at $\sim2\times10^6$ K.  In the medium-band image, the
emission appears confined to the edges of the soft-band
nebulosity, with a clump of brighter emission at the extreme
northern edge of the X-ray-emitting region.

\subsection{X-ray Surface Brightness vs.\ Optical/IR
Extinction}

Assuming Case B recombination, the hydrogen emission line ratios
$I_{H\alpha}/I_{P\alpha}$ and $I_{Br\gamma}/I_{Br\alpha}$
are relatively insensitive to temperature
in the regime of interest for planetary nebulae, $T
\sim 10^4$ K (e.g., Osterbrock 1989, Table 4.2). Therefore, we can
assume that deviations of the observed line ratios from
their ``canonical'' Case B recombination values are due to
extinction. Hence, assuming a standard interstellar
dependence of extinction on wavelength (e.g., Osterbrock 1989, Table
7.2), we use the spatial distribution of the ratios
$I_{H\alpha}/I_{P\alpha}$ and $I_{Br\gamma}/I_{Br\alpha}$,
as derived from the ratios of images presented in \S 2.2, to infer visual
extinction ($A_V$) as a function of position within \BD\ (Fig.\ 11)
and NGC 7027 (Fig.\ 12), respectively.

Harrington et al.\ (1997) and Robberto et al.\ (1993)
employed techniques similar to the preceding, to
infer the spatial distribution of $A_V$ for \BD\ and NGC
7027, respectively. Their results are
qualitatively similar to those derived here. The
extinction across both nebulae can be quite large, however,
and use of near-infrared images provides a better ``lever
arm'' for deducing $A_V$ in very highly obscured
regions. For example, the Robberto
et al.\ extinction map of NGC
7027, constructed from optical hydrogen
recombination line imaging, is limited to regions
with $A_V < 5$, whereas the $A_V$ map constructed from
longer-wavelength hydrogen transitions 
available to COB (Fig.\ 12) remains sensitive to $A_V \sim 15$. 

The distributions of $A_V$ are similar for the two nebulae,
in that the largest values of extinction lie near the
nebular perimeters, and extinction ``holes'' are observed
toward the nebular interiors. Generally, however, we find
$A_V$ within NGC 7027 (for which $A_V$ lies largely between
the range $1.5 \stackrel{<}{\sim} A_V \stackrel{<}{\sim}
15$) to be a factor 3-4 larger than $A_V$ within \BD\ (for
which $0.5 \stackrel{<}{\sim} A_V \stackrel{<}{\sim}
5$). 
Also, the dark lane apparent in high-resolution optical
images of NGC 7027 (e.g. KVS, their Fig.\ 1) appears as an
enhancement of $A_V$ in Fig.\ 12. This is consistent with the
hypothesis (KVS) that \BD\ and NGC 7027 share a common
structure --- bipolar, with a dense equatorial region ---
but that NGC 7027 is viewed at intermediate inclination
angle whereas \BD\ is viewed nearly pole-on. 

Notably, the X-ray surface brightnesses of both nebulae are
strongly anticorrelated with the local value of
$A_V$. In each case, the peak of X-ray emission lies very near
the minimum in $A_V$, and little or no X-ray emission is
observed toward regions of 
highest $A_V$. There is a gradient of $A_V$ in interior regions of \BD\ 
(i.e., regions of the nebula interior to the bright shell 
seen in H recombination lines), such that 
there is greater extinction
toward the southwestern side of the central star (a result
independently obtained by Harrington et al.\ 1997), and
there is an extinction ``hole'' 1-2$''$ to the eastern side of the
central star. The X-ray emission, correspondingly, is much
brighter to the east than to the west of the star and,
furthermore, the contours of X-ray surface brightness
closely follow the $A_V$ distribution in detail (Fig.\ 11). In NGC
7027, the relationship between $A_V$ and X-ray surface
brightness is also strong, with the lowest contours of
X-ray emission very closely tracing the regions of large $A_V$
(Fig.\ 12). In particular, the ``pinched waist'' apparent in the X-ray
emission morphology lies at the position of the
enhancement of $A_V$ that appears to define the equatorial
plane of the nebula. Also, the strongest X-ray emission from
NGC 7027 lies very near --- though slightly farther from the
central star than --- the extinction ``hole'' apparent to the
northwest of the star.

We have also constructed an H$\alpha$/H$\beta$ line ratio
map of NGC 6543 (not shown), from image data available in
the HST archive. In stark constrast to the line ratio maps
of \BD\ and NGC 7027, from which the $A_V$ maps in Figs.\
11, 12 and were constructed, the H {\sc ii} line ratio map
of NGC 6543 is smooth and featureless across its central,
X-ray-emitting region, indicating that any extinction of the
X-ray nebula is quite spatially uniform.  Furthermore, the
measured H$\alpha$/H$\beta$ ratio across this region,
$2.5\pm0.3$, agrees within uncertainty with the value
expected for Case B recombination, $\sim2.8$ (Osterbrock
1989). This result is consistent with the very small
inferred extinction toward NGC 6543 as measured by Tylenda
et al.\ (1992) and with the observation that X-ray absorption becomes
detectable by CXO only for energies $< 0.4$ keV (Chu et al.\ 2001).

\section{Discussion}

The comparisons of X-ray and extinction images in the
preceding section reveal striking correspondences between
the spatial distributions of $A_V$ and 
the X-ray surface brightnesses of \BD\ and NGC 7027.
These results suggest that intranebular extinction plays a very 
important role in determining the X-ray emission
morphologies of young, dusty PNs. 
The close correspondence of
regions of low extinction and bright X-ray emission in both
\BD\ and NGC 7027 suggests that some extended X-ray
emission may remain undetected, making it difficult to draw
conclusions as to the intrinsic shape (e.g., axisymmetric
vs.\ elliptically symmetric) of their soft X-ray emitting regions.

Extinction alone cannot fully explain the strong departures
from spherical symmetry and small-scale clumping of the
X-ray emission from all three nebulae, however. If
extinction were the sole contributor to the X-ray
asymmetries of \BD\ and NGC 7027, then we would expect these
asymmetries to be more extreme at lower energy, where
photons are easily absorbed, than at higher energy, where
the X-rays are more highly penetrating. By way of analogy,
we note that the near-infrared emission-line morphologies of
both nebulae are substantially more symmetric and regular
than their optical emission-line morphologies, a difference
that can be ascribed to the large decrease in the
probability of scattering or absorption of photons by
intranebular dust in the near-infrared, relative to the
optical regime.

In the X-ray imagery presented here, however, only NGC 7027
seems to display the expected trend of increasing symmetry
with increasing energy --- although even at high energy, its
X-ray morphology differs sharply from its shell-like
appearance in near-infrared emission-line images (Fig.\
7). \BD\ is observed to become clumpier and more one-sided
as X-ray energy increases. The hard-band emission from NGC
6543, for which extinction effects are negligible in the
bands considered here, is mostly confined to a small region
of the north lobe.  It seems, therefore, that strongly
position-dependent variations in the physical conditions in
the shocked gas are responsible, at least in part, for the
asymmetric and/or clumpy X-ray-emitting regions of these
PNs. Indeed, while it seems likely that intranebular
extinction modulates the nebular X-ray surface brightnesses
of \BD\ and NGC 7027, we cannot as yet rule out other
interpretations. For example, the regions of brightest X-ray
emission may correspond to the smallest $A_V$ because the dust
in these regions is being destroyed by the intense, high-energy
radiation.

Soker \& Kastner (2003) argue that the X-ray luminosities
and temperatures of PNs can be explained by a model in which
the diffuse emission is produced by a shocked, moderate
speed ($\sim500$ km s$^{-1}$) post-AGB wind, or a wind
emanating from a companion to the central post-AGB star,
where the shocked gas has had time to expand and cool
adiabatically. This wind and the distribution of previously
ejected AGB material also should largely determine the spatial
distribution of the X-ray emitting gas. The post-AGB or
companion wind may be highly collimated, resulting in strong
axial or point symmetries observable in X-rays. Other
processes might then govern the 
temperature structure (and therefore spatial distribution) of 
X-ray-emitting gas in detail, with heat conduction from 
shock-heated to photoionized gas in the presence of magnetic
fields and/or direct mixing of shock-heated and
photoionized gas being the most likely candidates
(Soker \& Kastner 2003, and references therein). These
temperature-moderating mechanisms may be more 
or less important in different nebulae. 

The images presented here, combined with the global X-ray
properties of PNs displaying diffuse X-ray emission (Soker
\& Kastner 2003, their Table 1), support this general
model. That is, adiabatic cooling of a shocked,
moderate-speed, post-AGB or companion wind should produce
X-ray luminosities and temperatures in the general range
observed ($\sim10^{32}$ ergs s$^{-1}$ and $1-3\times10^6$ K,
respectively). The detailed X-ray
surface brightness distributions of NGC 6543 and NGC
7027 suggest, moreover, that collimated outflows are
responsible for the detailed structure of their
X-ray emitting regions. 

In NGC 7027, which exhibits collimated
outflows in near-infrared line emission (Cox et al.\ 2002),
the action of such a highly directed (bipolar) fast wind
from the central star --- or, more likely, an as-yet
undetected companion\footnote{It seems unlikely that a
moderate-speed ($\sim 400$ km s$^{-1}$), collimated wind
could be driven by the present central star, as such a wind
speed is much smaller than the escape velocity from a star
of mass of $0.7 M_\odot$ and radius $0.07 R_\odot$ (Latter
et al.\ 2000b).} --- would readily explain its axisymmetric
appearance in the hard-band
image (Fig.\ 9) and the apparent ``breakout'' of
X-ray-emitting plasma beyond the bright rim of near-infrared
emission from ionized gas (Fig.\ 7). 
Indeed, the morphology of high-velocity
Br$\gamma$ emission is very similar to that of the X-ray
emission (Cox et al.). Similarly, both the optical emission
and the diffuse, soft X-ray emission from NGC 6543 appear
to be point-symmetric (see Fig.\ 10 and
Chu et al.\ 2001), suggestive of the action of
collimated outflows (although Chu et al.\ ascribe the
limb-brightened appearance of the X-ray emission from NGC 6543
to the effects of heat conduction or mixing of shocked plasma
with photoionized gas).

\BD\ may be an example of a PN in which heat conduction
and/or mixing play a crucial role in determining X-ray
morphology. Either mechanism would explain the bright,
compact X-ray-emitting region 
at the eastern rim of the nebula, with heat conduction
more naturally explaining its plasma abundances 
(Soker \& Kastner 2003).  However,
\BD\ and NGC 7027 bear strong resemblances in
many fundamental respects (e.g., Soker \& Kastner 2003) and,
in particular, \BD\ is also known to harbor 
collimated outflows (Bachiller et al.\ 2000). Hence the
asymmetric appearance of \BD\ at high energy, in comparison
with NGC 7027, may be primarily the result of the different
line-of-sight inclinations of these two nebulae (Masson
1989; KSVD; see, however, Bryce \& Mellema 1999, who find
that \BD\ more closely resembles the elliptical PN NGC 40). It remains
to demonstrate, however, how collimated outflows might produce the
off-center X-ray bright spot in \BD, as the outflows detected by Bachiller
et al.\ do not correspond spatially to this region of X-ray emission.

\section{Summary and Conclusions}

We have presented and analyzed reprocessed X-ray Chandra
X-ray Observatory images of the planetary nebulae \BD, NGC
7027, and NGC 6543. To maximize the spatial resolution of
the images, we employ an event relocation technique that
takes full advantage of CXO's exceptional spatial
resolution; we estimate that the reconstructed images
represent a $\sim30$\% improvement in PSF FWHM,
over the original images. Energy-resolved images demonstrate that the
morphologies of all three nebulae depend sensitively on
energy. Whereas NGC 7027 appears clumpiest at low
energy and more symmetric at higher energy, \BD\ becomes more
one-sided and asymmetric with increasing X-ray
energy. Diffuse emission from NGC 6543 is dominated by soft
X-rays, with its central point source dominant in the harder
energy bands.

With the aid of high-resolution optical and near-infrared H
{\sc ii} emission line images, we consider the extent to
which nonuniform intranebular absorption affects the X-ray
morphologies of these PNs. For both \BD\ and NGC 7027, we
find a strong anticorrelation between broad-band X-ray
surface brightness and the degree of visual extinction,
where the latter is inferred from the spatial distribution
of H {\sc ii} line ratios. We conclude that extinction plays
an important role in the overall X-ray appearances of these
young, dusty, molecule-rich PNs. The PN NGC 6543 has a much
smaller and more spatially uniform degree of visual
extinction than either \BD\ or NGC 7027, however, so
spatially variable absorption does not affect its X-ray morphology.

While extinction is important in determining the X-ray surface
brightnesses of \BD\ and NGC 7027, extinction alone cannot explain
many important characteristics of the X-ray morphologies of these
PNs. In particular, the highly asymmetric appearance of \BD\ and the
axisymmetric or point-symmetric appearances of NGC 7027 and NGC 6543
appear to point out intrinsic X-ray emission properties of these
nebulae. We conclude that the action of collimated outflows likely
dictates the X-ray appearances of at least two of three nebulae
studied here, with heat conduction in the presence of magnetic fields
potentially playing a more important role in the asymmetric surface
brightness of \BD.

Additional spatially resolved CXO/ACIS and XMM-Newton
observations of planetaries are required to better establish
the relative importance of the various processes --- e.g.,
collimated outflows, heat conduction, mixing of shock-heated
and photoionized plasma --- proposed to explain their X-ray
morphologies as well as their X-ray luminosities and
temperatures. We have demonstrated here that these
observations must be accompanied by subarcsecond
emission-line imaging in the optical and near-infrared, to
unambiguously distinguish between intrinsic and extrinsic
processes that can determine the energy-dependent X-ray
morphologies of PNs.

\acknowledgements{ The authors wish to thank two anonymous
referees for many helpful comments and suggestions.
J.H.K. and J.L. acknowledge support for this research
provided by NASA/CXO grant GO0--1067X to RIT. 
N.S. acknowledges support from the US-Israel
Binational Science Foundation. }



\begin{table}
\caption{Event Grade Distributions for Planetary Nebulae
Observed by CXO}
\begin{tabular}{ccrrrrrr}
\hline
 & & \multicolumn{2}{c}{NGC 6543} & \multicolumn{2}{c}{\BD}
 & \multicolumn{2}{c}{NGC 7027} \\
Event type & FLTGRADE & No. & \% & No. & \% & No. & \% \\
\hline
\hline
All     &  (all)   & 1755 & ...     & 4868 & ...    & 306 & ... \\
Single-pixel & 0   & 830  &  47.3 & 1934 & 39.7 &  
  95 & 31.0 \\
Two-pixel    & 2,8,16,64 & 764 & 43.5 & 2233 & 45.9 &
 161 & 52.6 \\
Corner-split & 10,11,18,22, & 127 & 7.2 & 456 & 9.4 & 40 &
 13.1 \\
             & 72,80,104,208 \\
Selected & 0,2,8,10,11, & 1721 & 98.1 & 4623 & 95.0 & 296 &
 96.7 \\
         & 16,18,22,64, \\
         & 72,80,104,208 \\
\hline
\end{tabular}
\end{table}

\singlespace

\begin{figure}[htb]
\includegraphics[scale=.8,angle=0]{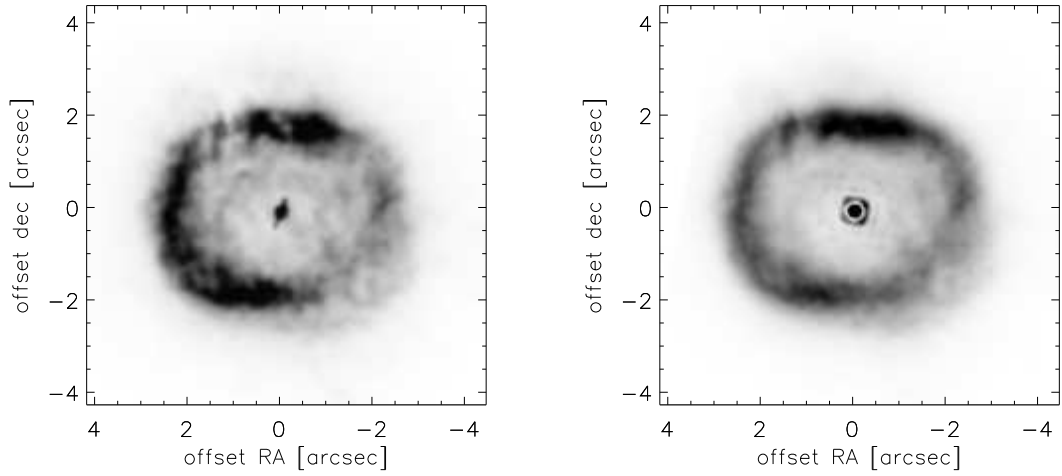}
\caption{{\it Hubble Space Telescope} H$\alpha$ 0.6563
$\mu$m (left) and P$\alpha$ 1.87 $\mu$m (right) images of
\BD\ obtained with WFPC2 and NICMOS, respectively. }
\end{figure}

\begin{figure}[htb]
\includegraphics[scale=.8,angle=0]{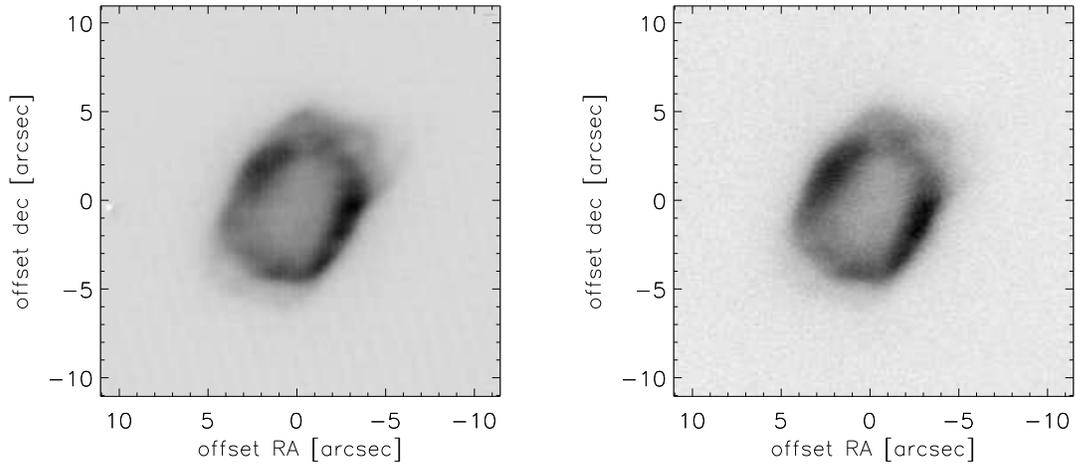}
\caption{Br$\gamma$ 2.16 $\mu$m (left) and Br$\alpha$
4.05 $\mu$m (right) images of NGC 7027, obtained with the
Cryogenic Optical Bench on the Kitt Peak 4 m telescope.}
\end{figure}

\begin{figure}[htb]
\includegraphics[scale=1,angle=0]{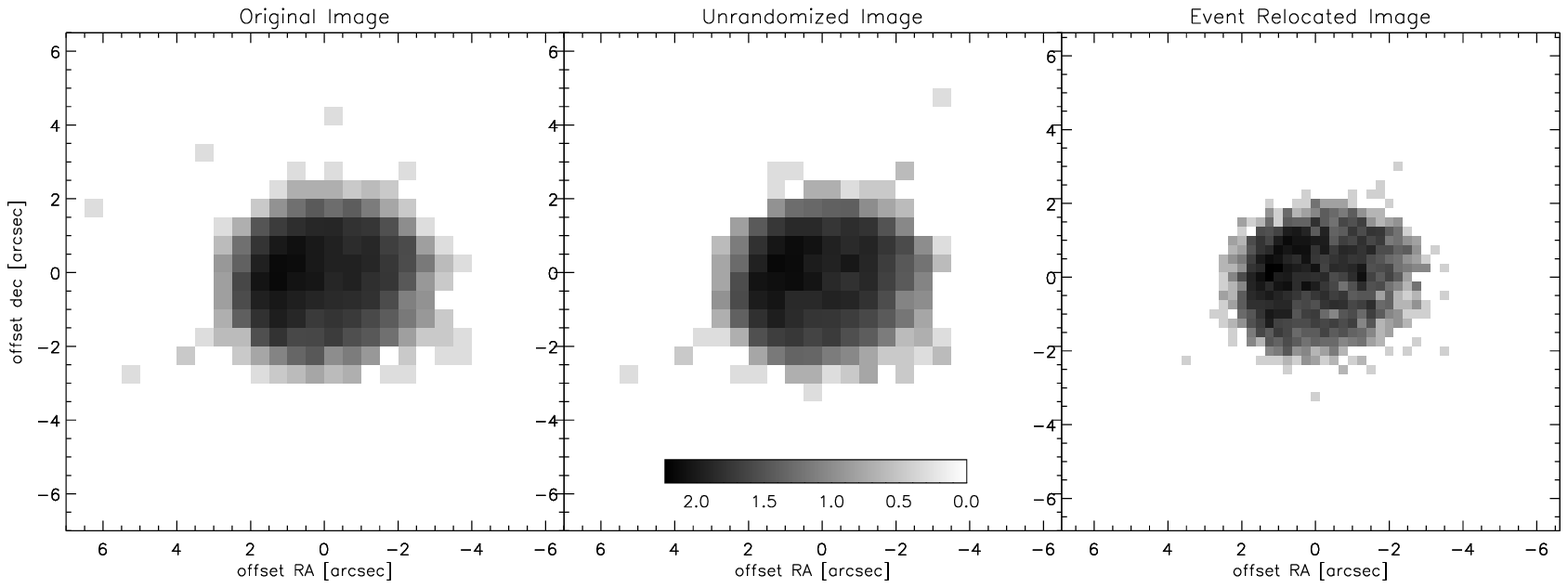}
\caption{Left: CXO X-ray image of \BD, obtained by binning
events before removing position randomization or applying sub-pixel
event position corrections. Center: image of \BD\ obtained by binning
events after removing event position 
randomization. Right: image of \BD\ obtained by binning
events after removing randomization and applying sub-pixel
event position corrections. Images in this and the next two
figures are presented in a log scale, with the bars
representing the greyscale mapping to log intensity.}
\end{figure}

\begin{figure}[htb]
\includegraphics[scale=1,angle=0]{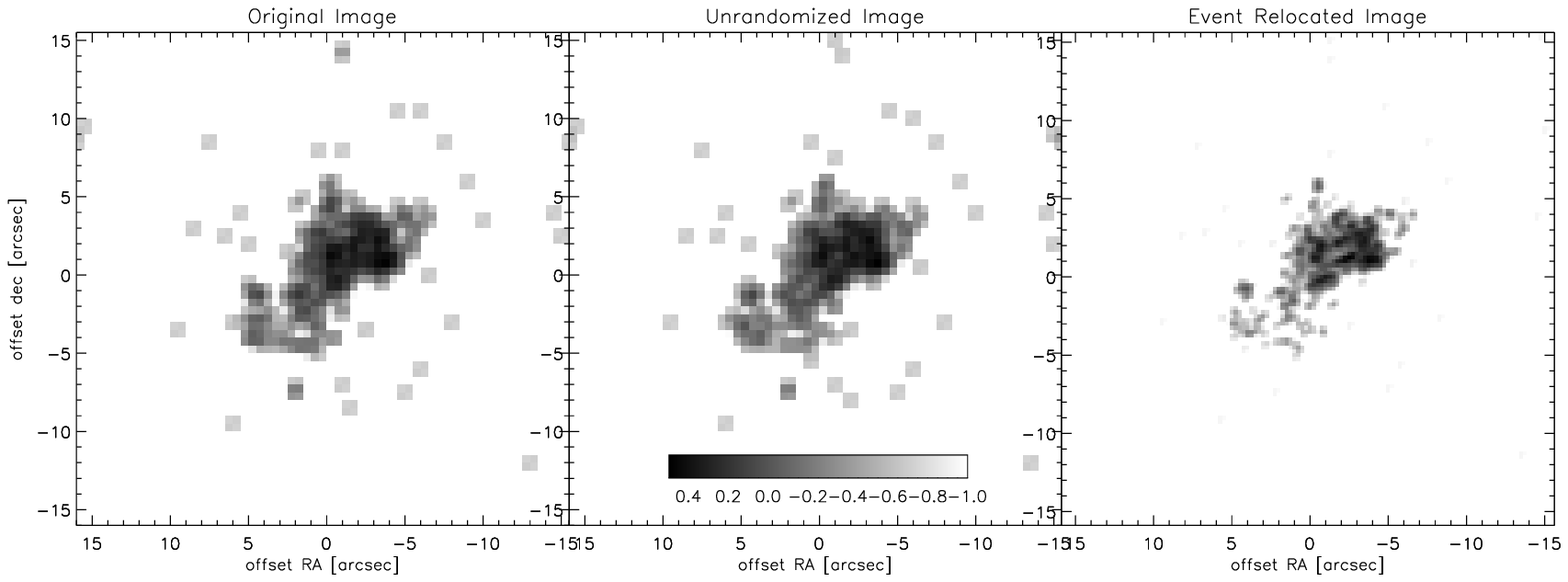}
\caption{Left: CXO X-ray image of NGC 7027, obtained by binning
events before removing position randomization or applying sub-pixel
event position corrections. Center: image of NGC 7027 obtained by binning
events after removing event position 
randomization. Right: image of NGC 7027 obtained by binning
events after removing randomization and applying sub-pixel
event position corrections. }
\end{figure}

\begin{figure}[htb]
\includegraphics[scale=1,angle=0]{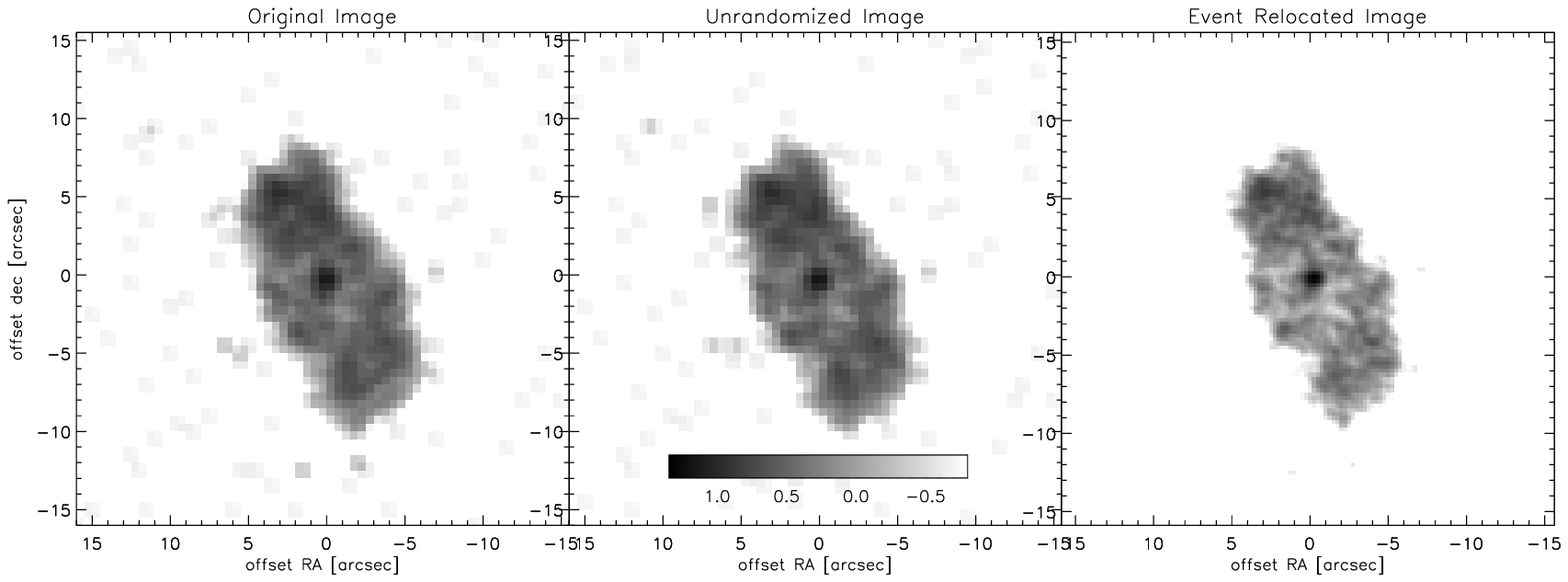}
\caption{Left: CXO X-ray image of NGC 6543, obtained by binning
events before removing position randomization and applying sub-pixel
event position corrections. Center: image of NGC 6543 obtained by binning
events after removing event position 
randomization. Right: image of NGC 6543 obtained by binning
events after removing randomization and applying sub-pixel
event position corrections. }
\end{figure}

\begin{figure}[htb]
\includegraphics[scale=1,angle=0]{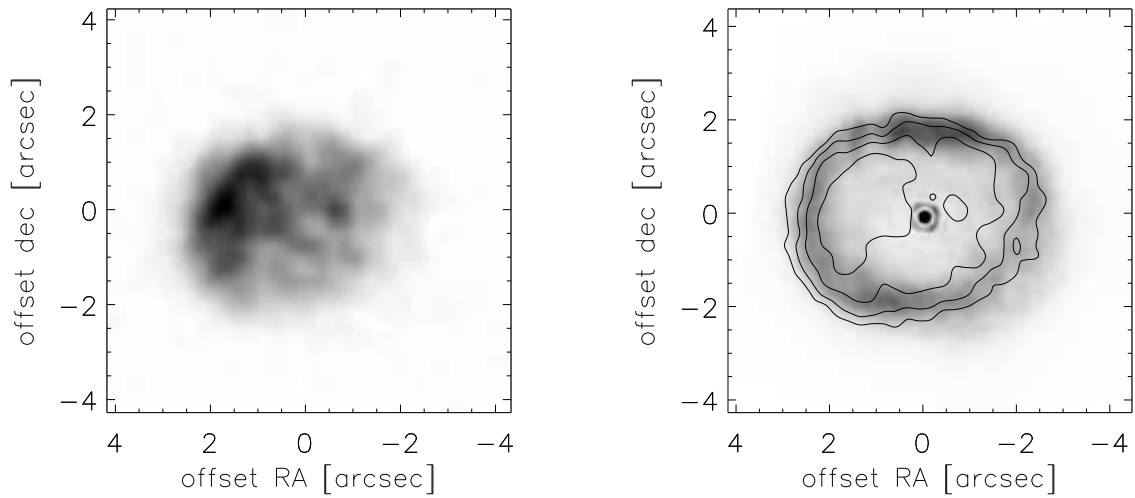}
\caption{Left: Reconstructed, broad-band CXO X-ray image of
\BD. Right: Contours of 
X-ray surface brightness 
overlaid on the P$\alpha$ image of Fig.\ 1. Contour levels
are at 12, 24, 48, and 96 counts arcsec$^{-2}$.
In this Figure and in Fig. 7,
the CXO image has been 
convolved with a Gaussian function with width
approximating that of the instrumental point spread
function. }
\end{figure}

\begin{figure}[htb]
\includegraphics[scale=1,angle=0]{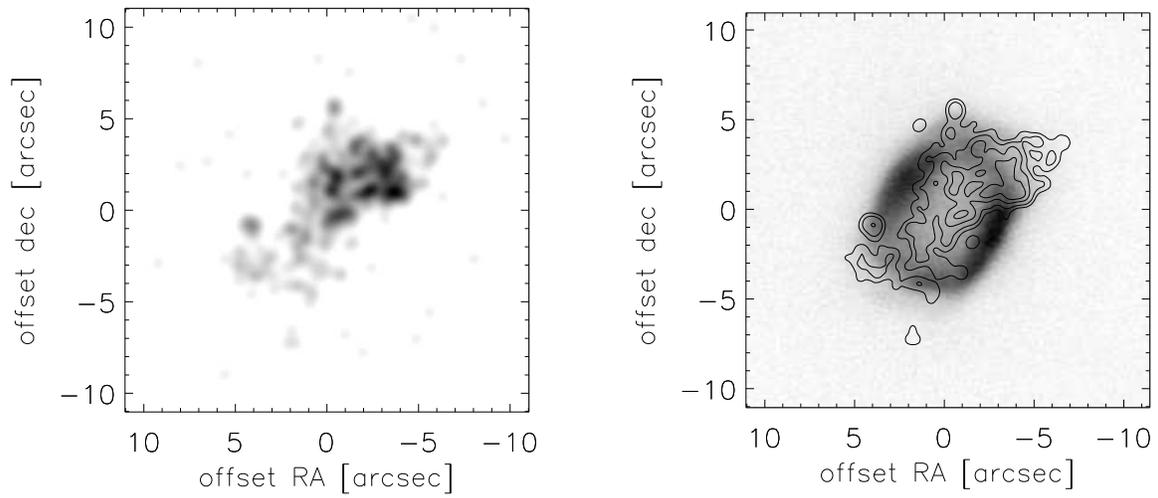}
\caption{Left: Reconstructed, broad-band CXO X-ray image of
NGC 7027. Right: Contours 
of X-ray surface brightness 
overlaid on the Br$\alpha$ image of Fig.\ 2. Contour levels
are at 4, 8, 16, and 24 counts arcsec$^{-2}$. In each
panel, offset (0,0) corresponds to the position of the
central star to $\sim0.2''$ (as determined from comparison
of the Br$\alpha$ image 
with HST near-infrared imagery; Latter et al.\ 2000b).} 
\end{figure}

\begin{figure}[htb]
\includegraphics[scale=1.,angle=0]{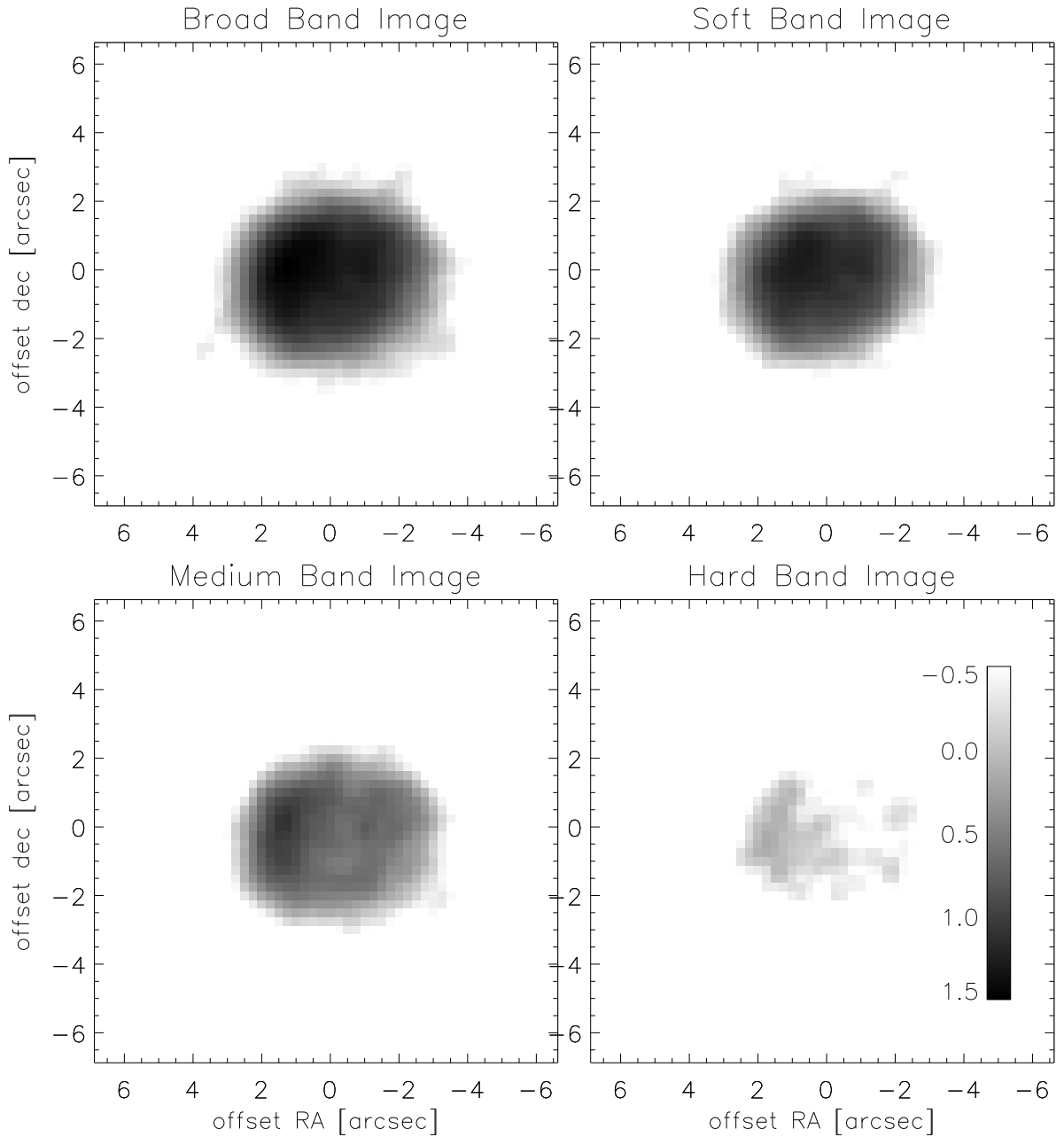}
\caption{Energy-resolved images of \BD. Upper left:
broad-band image, obtained from all events with energies 
$\le 3.0$ keV. Upper right: soft-band image, obtained from
events with energies $\le 0.7$ keV. Lower left: medium-band
image, obtained from events with energies between 0.7
keV and 1.2 keV. Lower right: hard-band image, obtained from 
events with energies between 1.2 keV and 3.0 keV. Images in
this and the next two 
figures are presented in a log scale, with the bars
representing the greyscale mapping to log intensity.}
\end{figure}

\begin{figure}[htb]
\includegraphics[scale=1.,angle=0]{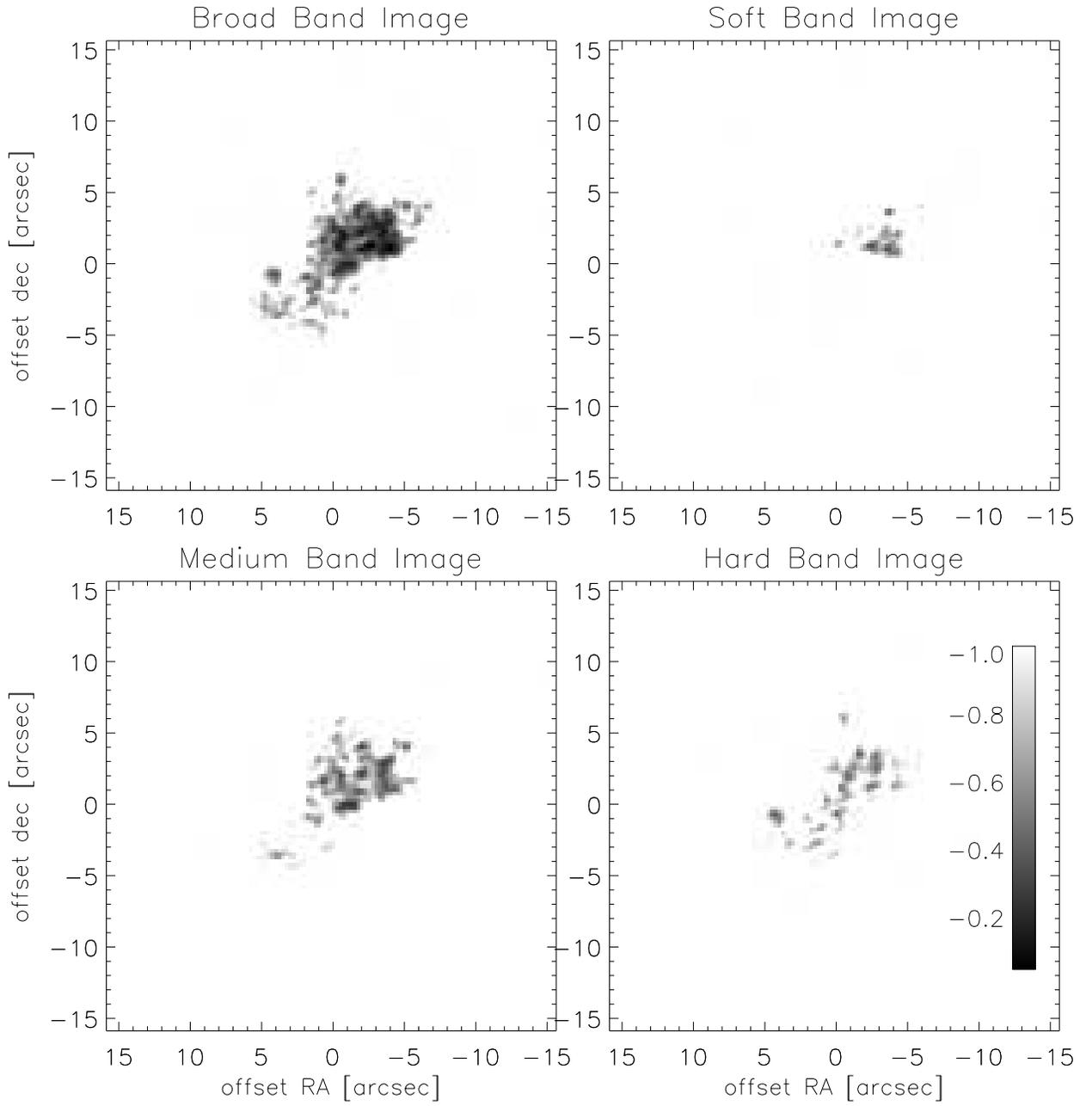}
\caption{Energy-resolved images of NGC 7027. 
Energy ranges for panels are as in the previous figure.}
\end{figure}

\begin{figure}[htb]
\includegraphics[scale=1.,angle=0]{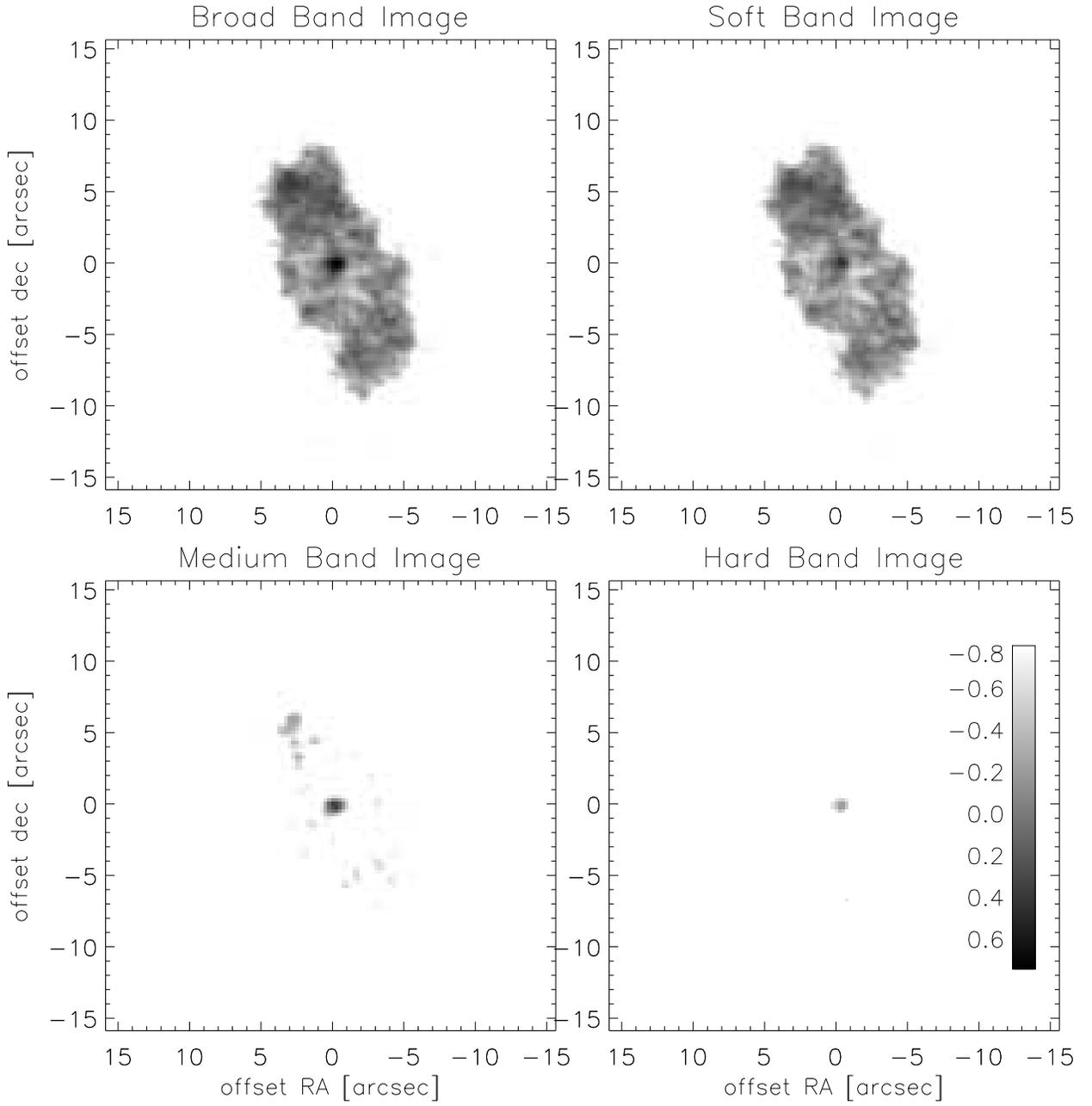}
\caption{Energy-resolved images of NGC 6543. 
Energy ranges for panels are as in Fig.\ 8.}
\end{figure}

\begin{figure}[htb]
\includegraphics[scale=1.,angle=0]{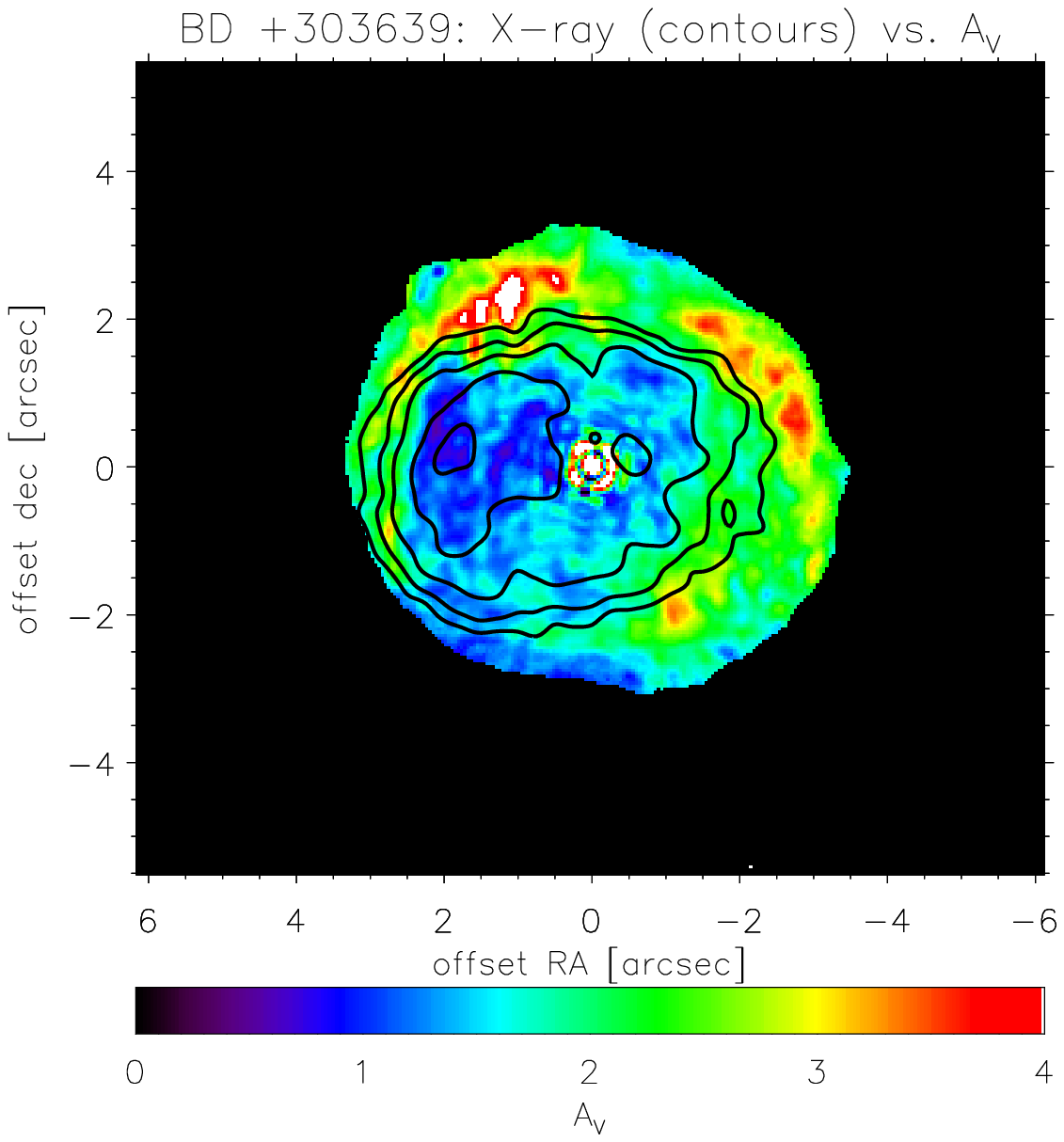}
\caption{Overlay of CXO X-ray image (contours) on
extinction map of \BD\ (color).}
\end{figure}

\begin{figure}[htb]
\includegraphics[scale=1.,angle=0]{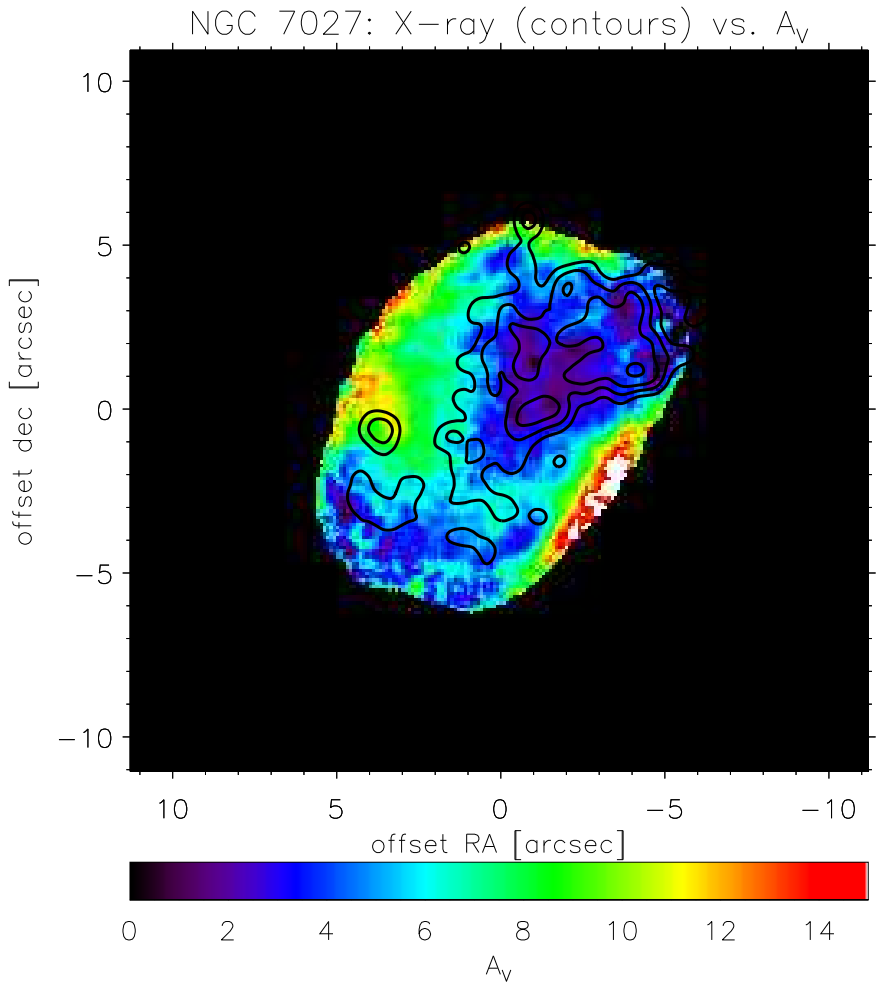}
\caption{Overlay of CXO X-ray image (contours) on
extinction map of NGC 7027 (color).}
\end{figure}


\end{document}